\documentclass[sigconf]{acmart}
\usepackage[suppress]{color-edits}
\usepackage{pdflscape}
\addauthor{bk}{blue}
\addauthor{hh}{red}
\addauthor{rk}{cyan}

\usepackage{booktabs}

\newcommand{\xhdr}[1]{\vspace{1.7mm}\noindent{{\bf #1.}}}

\copyrightyear{2025}
\acmYear{2025}
\setcopyright{cc}
\setcctype{by}
\acmConference[FAccT '25]{The 2025 ACM Conference on Fairness, Accountability, and Transparency}{June 23--26, 2025}{Athens, Greece}
\acmBooktitle{The 2025 ACM Conference on Fairness, Accountability, and Transparency (FAccT '25), June 23--26, 2025, Athens, Greece}\acmDOI{10.1145/3715275.3732191}
\acmISBN{979-8-4007-1482-5/2025/06}

\begin{document}

% Keywords: Reponsible AI tools Meta-review  Stages Roles/Actors Uneven distribution
\title[A Systematic Review and Meta-Analysis of (Actor, Stage)-Specific Tools for Responsible AI]{The ``Who'', ``What'', and ``How'' of Responsible AI Governance: \\A Systematic Review and Meta-Analysis of (Actor, Stage)-Specific Tools}

\author{Blaine Kuehnert}
\orcid{0009-0001-5702-5693}
\authornote{Both authors contributed equally to this research.}
\affiliation{%
  \institution{Carnegie Mellon University}
  \city{Pittsburgh}
  \state{Pennsylvania}
  \country{USA}
}
\email{blainekuehnert@cmu.edu}

\author{Rachel M. Kim}
\orcid{0009-0000-3325-243X}
\authornotemark[1]
\affiliation{%
  \institution{Carnegie Mellon University}
  \city{Pittsburgh}
  \state{Pennsylvania}
  \country{USA}
}
\email{rachelmkim@cmu.edu}

\author{Jodi Forlizzi}
\orcid{0000-0002-7161-075X}
\affiliation{%
  \institution{Carnegie Mellon University}
  \city{Pittsburgh}
  \state{Pennsylvania}
  \country{USA}
}
\email{forlizzi@andrew.cmu.edu}

\author{Hoda Heidari}
\orcid{0000-0003-3710-4076}
\affiliation{%
  \institution{Carnegie Mellon University}
  \city{Pittsburgh}
  \state{Pennsylvania}
  \country{USA}
}
\email{hheidari@andrew.cmu.edu}

\renewcommand{\shortauthors}{Kuehnert et al.}

\begin{abstract}
  The implementation of responsible \rkedit{Artificial Intelligence} (AI) in an organization is inherently complex due to the involvement of multiple stakeholders, each with their unique set of goals and responsibilities across the entire AI lifecycle. These responsibilities are often ambiguously defined and assigned, leading to confusion, miscommunication, and inefficiencies. Even when responsibilities are clearly defined and assigned to specific roles, the corresponding AI actors lack effective tools to support their execution. 
  Toward closing these gaps, we present a systematic review and comprehensive meta-analysis of the current state of responsible AI tools, focusing on their alignment with specific stakeholder roles and their responsibilities in various AI lifecycle stages. We categorize over 220 tools according to AI actors and stages they address. Our findings reveal significant imbalances across the stakeholder roles and lifecycle stages addressed. The vast majority of available tools have been created to support AI designers and developers specifically during data-centric and statistical modeling stages while neglecting other roles such as \rkedit{organizational} leaders, deployers, end-users, and impacted communities, and stages such as value proposition and deployment. \rkedit{This} uneven distribution highlights critical gaps that currently exist in responsible AI governance research and practice. Our analysis reveals that despite the myriad of frameworks and tools for responsible AI, it remains unclear \emph{who} within an organization and \emph{when} in the AI lifecycle a tool applies. Furthermore, existing tools are rarely validated, leaving critical gaps in their usability and effectiveness. These gaps provide a starting point for researchers and practitioners to create more effective and holistic approaches to responsible AI development and governance. 
  %Our work is also a call for organizational leaders to include a variety of stakeholders across the AI lifecycle for better, more appropriate AI products. 
\end{abstract}




\maketitle

\section{Introduction}
As the use of Artificial Intelligence (AI) to automate or support critical tasks proliferates, a wide range of interested parties have created principles, frameworks, and tools to ensure that AI systems are aligned with stakeholders' values.
In spite of the growing amount of interest in Responsible AI (RAI) and AI governance, we continue to see AI systems produce unintended outputs and fail to adhere to ethical principles.
For example, researchers have found that AI tools to detect COVID-19 have been ineffective, and potentially even harmful~\cite{heaven2021hundreds}.
% https://www.technologyreview.com/2021/07/30/1030329/machine-learning-ai-failed-covid-hospital-diagnosis-pandemic/
Faulty facial recognition software has led to the wrongful arrest of three men in the US~\cite{johnson2022how}.
% https://www.wired.com/story/wrongful-arrests-ai-derailed-3-mens-lives/
New York City's AI chatbot has encouraged businesses to break the law~\cite{lecher2024nyc}.
% https://themarkup.org/news/2024/03/29/nycs-ai-chatbot-tells-businesses-to-break-the-law

One barrier to translating the principles, frameworks, and tools to practice is that many are too general, focusing on high-level ethical principles such as autonomy and justice, but providing little guidance on how to promote the principles in action~\cite{Morley2019, schiff2020principles}.
On the other hand, some tools are too granular. For example, exclusively focusing on one technical definition of fairness during model testing does not necessarily guarantee fairness during AI system deployment~\cite{Kaye2023}.
Furthermore, many principles, frameworks, and tools have not been empirically tested for usability---the extent to which a RAI principle, framework, or tool is easy to adopt by practitioners---nor effectiveness---the extent to which using a principle, framework, or tool results in the claimed benefit in the real-world~\cite{mittelstadt2019ai, Kaye2023, berman2024scoping, cartwright2009thing}. 
% \hhcomment{We need to be clear what we mean by terms such as validation.}

More broadly, effective AI governance remains an inherently complex issue due to the numerous stakeholders and AI lifecycle stages involved.
First, AI systems have multiple stakeholders, including technical stakeholders (e.g., developers), business stakeholders (e.g., organizational leaders), and members of the general public.
These stakeholders differ in educational background, their approach to framing problems, and norms for communication~\cite{schiff2020principles}. Consequently, not all RAI frameworks and tools are usable by every stakeholder.
% Furthermore, the inclusion of numerous stakeholders in an AI system makes it difficult to clearly assign responsibilities, credit, or blame to a specific actor~\cite{schiff2020principles}.
% https://dl.acm.org/doi/10.1145/3531146.3533150
Second, preventing and mitigating AI harms requires understanding their source in the AI lifecycle---the process by which an AI model is imagined, designed, developed, evaluated, and integrated into broader decision-making processes and workflows. 
Thus, despite the growing collection of frameworks and tools for AI governance, \textbf{it remains unclear \emph{who} within an organization should conduct \emph{what} task at different stages of the AI lifecycle, and \emph{how}.} In particular, stakeholders are frequently unaware or uncertain about which principles, frameworks, and tools they can utilize to discharge their responsibilties. %\emph{concrete} and \emph{validated} framework or tool applies. 
This reality, in turn, makes it difficult for organizations to effectively govern AI. 

Past work has recognized this need for concrete RAI tools~\cite{Morley2019, Prem2023}.
There have been previous classifications of RAI tools by the stages of the AI life cycle where the tool is supposed to be used~\cite{Morley2019, Prem2023, Ayling2022} and, less commonly, by the stakeholders who can use a given tool and apply its output~\cite{Ayling2022}.
However, there is a lack of a \emph{comprehensive} review of current tools that classifies artifacts into the \emph{intersection} of stakeholders and AI lifecycle stages. This gap makes it difficult for each stakeholder in an AI system to know what tools are available to them at each stage of the AI lifecycle.
Furthermore, previous work does not check whether different artifacts were tested for usability or effectiveness.
This also makes it difficult for researchers and practitioners who develop AI governance tools to determine where additional research and development is needed.

\begin{figure*}
     \centering
     \includegraphics[width=1\linewidth]{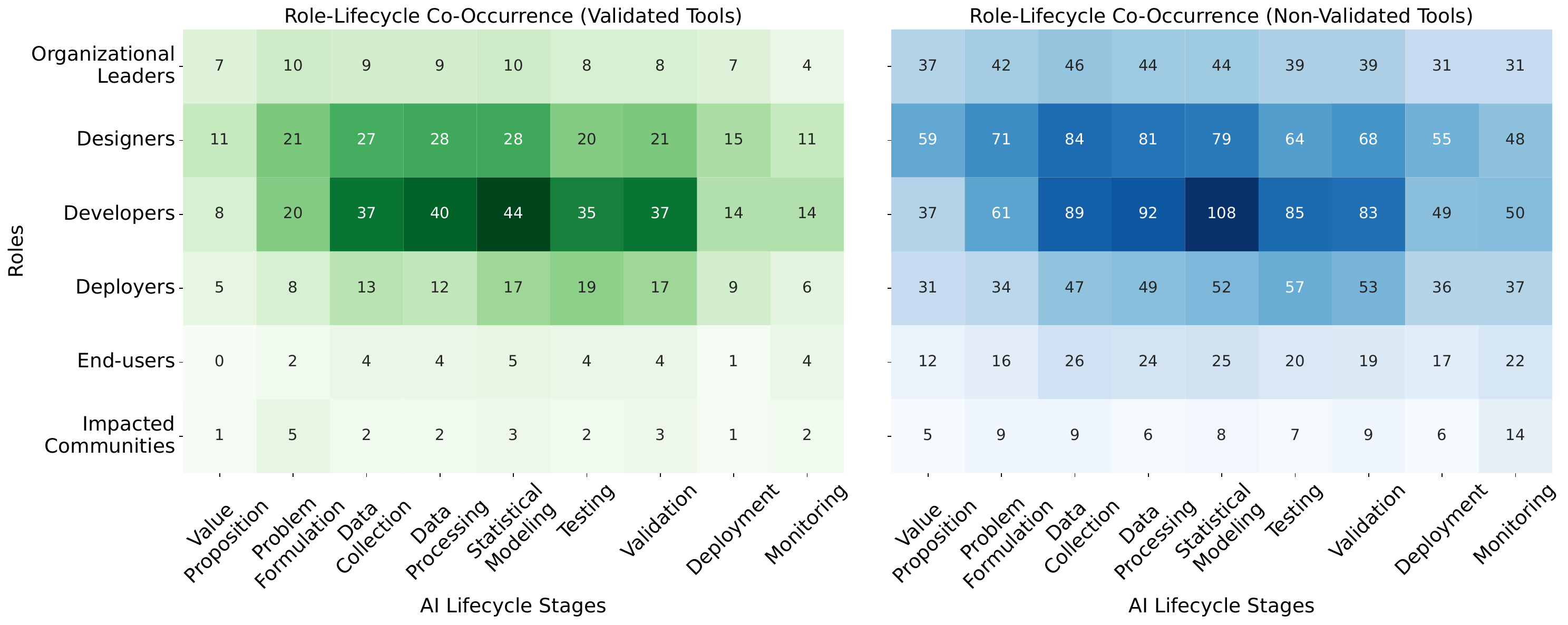}
     \caption{Validated \rkedit{co-occurrence matrix (left)} and \rkedit{validated and} non-validated \rkedit{co-occurrence matrix (right)} for stakeholders and stages}
     \label{fig:roles_stages_sbs}
\end{figure*}

% \xhdr{The Present Work} 
In this paper, we address the above gaps by conducting a systematic review of the literature of more than 220 AI governance and RAI tools that provide concrete guidance towards operationalizing RAI. We analyze these tools, categorizing them according to the specific stakeholder group and the AI lifecycle stage they address (Figure~\ref{fig:roles_stages_sbs}). 
We also assess whether each tool has been validated in any way \rkedit{prior to the release of the tool}. \rkedit{We define ``validation'' as  empirical (or even suggestive) evidence of usability and efficacy in practice} (for example, through a case study, experiment, or a pilot program). The resulting \rkedit{(stakeholder, stage)-}matrices \rkedit{(Figure~\ref{fig:roles_stages_sbs})} provide an overview of available resources, and offers practitioners a valuable resource to consult in deciding how to comply with RAI values in different elements of their work. In addition, it provides the RAI research community an empirical account of the gaps in the field and outlines impactful avenues for future work.

% \hhcomment{We need to clearly state what we mean by tool vs. framework and use the terms consistently throughout.}
\textbf{A note on key concepts and terminology:} We differentiate RAI ``tools,'' ``artifacts,'' and ``processes'' from ``principles'' and ``frameworks'' in two main ways. First, we differentiate by functional scope. In this respect, frameworks are conceptual structures or overarching guidelines that often define principles, directives, and high-level strategies for RAI. On the other hand, tools are specific, practical instruments, techniques, or processes that implement or operationalize certain aspects of a framework. The second differentiation pertains to the level of the focus in each approach. Frameworks are abstract and intended to guide thinking and high-level strategizing about RAI. They often deal with the ``what'' and ``why'' perspective, focusing on desired principles. In contrast, tools are applied, hands-on resources that assist in the implementation of a framework's principles in practice. They deal with the ``how'' by offering concrete steps and methods to achieve the desired goals of a framework. For the rest of this study, we focus our efforts on tools as opposed to frameworks. 
% Henceforth, we use the term RAI ``tool,'' ``artifact,'' or ``process'' to denote concrete guidance that is neither too high-level or too granular in the ways we described above.
%\Rachelcomment{unsure if it's necessary to explain this much given that we talked about the principles to practices gap already. There's a shorter version commented out in the Overleaf, but it might be too succinct.}

% \begin{figure}[]
% \centering
% \includegraphics[width=\textwidth]{figs/roles_stages_validated.png}
% \caption{Co-occurrence of roles and stages for validated tools}
% \label{fig:roles_stages_validated}
% \end{figure}

% \begin{figure}[]
% \centering
% \begin{minipage}{.4\textwidth}
%   \centering
%   \includegraphics[width=\linewidth]{figs/roles_stages_validated.png}
%   \captionof{figure}{}
%   \label{fig:roles_stages_validated}
% \end{minipage}%
% \begin{minipage}{.4\textwidth}
%   \centering
%   \includegraphics[width=\linewidth]{figs/roles_stages.png}
%   \captionof{figure}{}
%   \label{fig:roles_stages}
% \end{minipage}
% \caption{Summarization and so what}
% \end{figure}

% \TODO{Summary of main findings, after we have a draft of the discussion.}
Through our analysis of existing tools, we find that existing RAI tools are primarily developed for designers and developers in the data collection, data processing, statistical modeling, and testing stages.
Concrete tools for leaders, end-users, and impacted communities, in value proposition and problem formulation are comparatively lacking. 
Additionally, only \rkedit{36.6}\% of the RAI tools in our literature search mentioned any evidence that the tool creators checked the tool's usability or effectiveness.
Moreover, there are very few validated tools that look end-to-end across the AI lifecycle and engage with end-users and impacted communities along with other stakeholder groups.
% \TODO{Rachel: rewrite a shorter summary of what the findings mean}
% Our findings surrounding the over-represented stakeholders and stages suggest that current approaches to AI governance are mostly technical.
% Furthermore, the under-representation of leaders, end-users, and impacted communities suggests that current efforts toward effective AI governance underestimate the role of organizational culture and incentives, end-user experiences, and downstream consequences on ensuring responsible AI. 
% The relative lack of tools for value proposition and problem formulation indicates that responsible AI is often seen as an afterthought; ethical values are usually embedded after an AI system is created, failing to consider whether it is beneficial or even moral to create a AI system for a particular use case in the first place.
% This finding suggests the need for \emph{proven} tools for AI governance.
% \hhcomment{The following is your positive contribution. You should lead with this.}
% \Rachelcomment{Is this referring to the sentence that begins with ``overall''?}
%In summary, our systematic literature review provides stakeholders with a view of the tools available to them at each stage of the AI lifecycle, which tools are validated, and the broader RAI community with recommendations to help ensure responsible AI in practice.
We argue that without validated RAI tools, many tools may be improperly used and create a false sense of assurance. 
Furthermore, without tools that provide a holistic view across AI lifecycle stages and stakeholders, organizations risk employing a fragmented approach to AI governance. 
In line with these problems, we make three recommendations for practitioners and researchers: validating existing and new RAI tools, approaching AI governance holistically, and using the stakeholder-stage matrix as a blueprint for developing a custom approach to RAI depending on organizational and stakeholders' needs.

\section{Background and Related Work}
% - Overview of existing AI governance frameworks, approaches, philosophies.\\
% - Discussion of related methodologies for assessing governance practices.\\
% - Prior attempts to address roles or lifecycle stages individually.\\
% - Positioning this work in the context of FAccT’s mission: Create best practices for AI ethics; Advance research and dialogue on creating fair, accountable, and transparent systems; Foster collaboration between data and computer scientists, community organizers, advocates, and policy actors; Address issues related to fairness, justice, accountability, transparency, ethics, and adverse impacts of computational systems.

Our work builds upon previous literature reviews and collections of responsible AI frameworks and tools. This section highlights previous work by government agencies, organizations, and researchers to develop responsible AI frameworks and tools, past reviews and collections of these artifacts, \rkedit{and existing RAI tool validations}.

\xhdr{Existing Responsible AI Frameworks and Tools}
The rise of artificial intelligence (AI) in high-stakes domains has led to multiple approaches designed to address responsible AI practices.
Despite the growing body of work surrounding such principles, frameworks, and tools, many challenges remain in ensuring responsible AI \emph{in practice}. 
One challenge is that many guidelines, such as the European Commission's Ethical Guidelines for Trustworthy AI~\cite{highlevel2019ethics}, are too general~\cite{floridi2021establishing}.
There is a lack of methods that translate high-level principles into low-level requirements~\cite{mittelstadt2019ai}.
In fact, 78\% of technology workers wanted more guidance on dealing with ethical issues~\cite{doteveryone2019people}; relatedly, past work has called for moving from the ``what'' to the ``how'' of AI ethics~\cite{Morley2019}.

A set of tools has been developed with the goal of providing very concrete and low-level guidance, such as the four-fifths rule~\cite{us2023select}, Local Interpretable Model-agnostic Explanations (LIME)~\cite{lime}, and Shapley Additive Explanations (SHAP)~\cite{lundberg2017unified}.
However, there are concerns surrounding the narrow focus of the tools in terms of the AI lifecycle stages covered~\cite{Kaye2023, Ojewale2024}, type of methodology used~\cite{Ojewale2024}, and integrity in achieving the higher-level ethical principles~\cite{Ojewale2024}.
These concerns have highlighted the need for AI governance frameworks and tools to be both broad and flexible~\cite{Ojewale2024, schiff2020principles}.

AI governance principles, frameworks, and tools should also be validated for usability and effectiveness.
However, empirically proven methods for responsible AI are lacking~\cite{mittelstadt2019ai}.
Many AI auditing tools do not obviously consult practitioners or affected stakeholders~\cite{Ojewale2024}.
Another big limitation of existing principles, frameworks, and tools is the lack of knowledge surrounding when a certain artifact is appropriate and when it is not~\cite{Kaye2023}.
Consequently, there are calls for principle, framework, and tool creators to follow a quality assurance process~\cite{Kaye2023}.

In our work, we focus on tools that provide concrete yet not overly granular guidance towards RAI, finding the artifacts that are most useful in achieving good AI governance \emph{in practice}. 
Moreover, we label each artifact with whether or not there was validation of the applicability and usability of the framework or tool.

\xhdr{Reviews and Collections of Responsible AI Frameworks and Tools}
There is prior research on collecting, reviewing, and analyzing responsible AI frameworks and tools for all stages of the AI development lifecycle.
Morley et al.~\cite{Morley2019} and Prem~\cite{Prem2023} focus specifically on tools that are meant for developers, 
while Ojewale et al.~\cite{Ojewale2024} focus on auditing.
Some work~\cite{Johnson2023} only provides a collection of existing frameworks and tools available to organizations developing AI systems;
other work~\cite{Ortega2024, Morley2019, Prem2023} classifies the frameworks and tools according to the ethical principles that are addressed, including autonomy, beneficence, explainability, justice, \rkedit{and} non-maleficence.
Prem~\cite{Prem2023} and Ojewale et al.~\cite{Ojewale2024} also label each framework and tool with their type: for example, whether a tool is designed to identify relevant standards.
More related to our paper, there have been classifications~\cite{Ayling2022, Ortega2024, Morley2019, Prem2023} of existing frameworks and tools into the AI lifecycle stages to which the framework or tool should be applied.
Perhaps most similar to our work is the work of Ayling and Chapman~\cite{Ayling2022}, who label 39 AI ethics frameworks and tools with the stages of the AI lifecycle when the framework should be applied, the stakeholders who apply the tool, and the stakeholders who use the tool output.

Our work expands on these reviews by conducting a systematic literature review; we reviewed all of the frameworks and tools explored in each of these previous literature reviews. 
We consider the intersection of stakeholders and lifecycle stages and ask what tools are available to each stakeholder that address which lifecycle stages. Different from previous work, we focus on the stages that are \emph{addressed} by a framework or a tool rather than the stages that the framework or tool should be \emph{used in}.
Through \rkedit{our literature review}, we aim to help various stakeholders across an organization determine exactly which practical tools are available to them at every stage of the lifecycle.

\xhdr{RAI Tool Validations}
%%% Rachel's rough notes
\rkedit{Past work has recognized the need to validate RAI tools and highlighted the risks of failing to validate tools, including tool misuse~\cite{lee2021landscape} and over-reliance~\cite{kaur2020interpreting, kumar2021shapley}.
Consequently, researchers have performed validation studies on existing tools and found gaps between RAI tool capabilities and real-world needs.
For example, Kumar et al.~\cite{kumar2020problems} find that the theoretical properties of SHAP values do not align with the motivations behind why people desire AI explainability.
Deng et al.~\cite{deng2022exploring} perform an in-depth empirical exploration of how industry practitioners try to use fairness toolkits and discover that AI practitioners desire tools with more use-specific and context-specific guidance and support in integrating the tools with existing practices.
Berman et al.~\cite{berman2024scoping} performed a scoping study of 37 publications that discuss RAI tool validation, finding that RAI tool validations often employ HCI methods such as workshops, semi-structured interviews, and think-aloud studies.}
% past work has already recognized the need for the validation of tools, misalignment with practices, etc.
% towards this end, researchers have performed validation studies on existing tools after they have been published
% key findings...
% \cite{deng2023exploring} -- 
% \cite{kumar2021shapley} -- 
% \cite{berman2024scoping} -- sampled their list of tools, didn't do an analysis on the entire set of RAI tools that they had
% \cite{berman2024scoping} -- perform a scoping study of RAI evaluation practices. they find that some common evaluation types include HCI methods, including workshops, worked examples, and surveys

\rkedit{In this paper, we assess each tool for evidence of validation \emph{within} the tool's documentation, done prior to publishing the tool.
Note that by restricting our scope in this manner, we potentially exclude much of the important post-hoc work done by researchers~\citep[e.g.][]{kumar2020problems, deng2022exploring, richardson2021towards}.
However, due to the previously-mentioned risks of using non-validated tools, we believe that it is critical to ensure that tools are validated prior to their public release.}
\section{Methodology}
\subsection{Literature Review}
\xhdr{Terminology Selection and Refinement} The literature selection began with a focused list of key terms chosen to cover a breadth of AI governance framework descriptions. These included common words and phrases such as AI / artificial intelligence, ML / machine learning, framework, responsible, governance, stakeholder. Using these terms, we conducted an initial search on Google Scholar and Scopus, yielding a limited but targeted set of papers. From this initial collection, we identified synonyms and related terms to expand the search. 
Additional terms included transparency, assurance, auditing, risk management, regulation, ethics / ethical. This expanded list was used to generate combinations of terms to comprehensively capture the relevant literature. 
Examples of combinations included ``AI framework'', ``responsible AI'', ``AI governance'', ``AI risk management'', ``responsible AI framework'', ``responsible  AI stakeholder'', ``responsible AI auditing'', ``AI assurance'', ``responsible AI governance'', ``ethical AI framework''.
% \Rachelcomment{Unsure how many examples of combinations we need. Personally I feel like less could suffice.}

These term combinations were applied two times, first using AI/artificial intelligence, then replacing the term with ML/machine learning. 
% to cover the broadest set of terms possible.
We then collected the top 100 papers from Scopus and Google Scholar for each combination. The process resulted in overlapping results, which were deduplicated by removing articles that matched either DOI number or title, resulting in 1313 unique papers.
Five of these papers~\cite{Ayling2022, Ortega2024, Prem2023, Ojewale2024, Morley2019} were also reviews of existing AI frameworks and tools; we worked through each of the frameworks and tools contained in these papers and added them to the initial list as well. 
\bkedit{These surveys also identified open source, industry, and other tools not published in academic domains. 
Upon reading the literature search methodology of these papers, we determined that the methodology in these papers was sound (for example, many of the surveys included a keyword search and selection criteria similar to ours in non-academic domains, such as English Google searches). 
Thus, we decided to rely on these sources to identify relevant work in non-academic settings.}
It is worth noting that this broad selection approach resulted in finding several papers that use reversed versions of the terms in specific technologies or use cases, for example using AI for financial auditing rather than processes or tools for auditing AI. These application-specific papers did not apply to our current research and were excluded from the final selection, reducing the number of applicable papers to 416. 

\xhdr{Inclusion and Exclusion Criteria}
To further refine the dataset, we assessed each paper for applicability using the inclusion and exclusion criteria found in Table~\ref{tab:inclusion_exclusion_criteria}.
After applying these criteria, we finalized a set of \rkedit{224} tools for this analysis\footnote{The final data selection was made available at the following link: 
\href{https://docs.google.com/spreadsheets/d/13012Km2BWUerniKiy87UeoOCpOf9MoCoIVMpNp-SKSA/edit?usp=sharing}{Click here}
}.

\bkedit{In addition to the search described above, we looked into existing catalogs of RAI tools published and maintained by entities including NIST and OECD and observed that, while these catalogs were extensive, nearly all of the entries did not meet the inclusion criteria and lacked the specific tools necessary for practical implementation.}

\begin{table*}[]
\scriptsize
    \centering
    \caption{Inclusion and exclusion criteria for the literature review.}
    \begin{tabular}{p{0.45\textwidth}p{0.45\textwidth}}
         \toprule
         \multicolumn{1}{c}{\textbf{Inclusion Criteria}} & \multicolumn{1}{c}{\textbf{Exclusion Criteria}} \\
         \midrule
\begin{itemize}
    \item Concrete and explicit references: We selected works that clearly stated the roles (e.g., developers, leaders) and/or lifecycle stages (e.g., data collection, model validation) addressed, for example, ``this tool is for developers to use during data collection.''
    \item Explicit methodology: We chose to focus on papers describing concrete implementation methods, such as code, questionnaires, case studies, or detailed guides.
    \item Clear responsible AI-related purpose: We selected works that were specifically designed and created for responsible AI.
    \item Language: We only selected works that were published in English.
\end{itemize}
         & 
\begin{itemize}
    \item Purely conceptual papers: We excluded works focusing solely on abstract principles or ethics without actionable insights.
    \item High-level guidance: We excluded works that relied on normative statements or recommendations lacking specific implementation details, for example, ``organizations should consider fairness.''
    \item Overly granular guidance: We excluded works that are too specific in focus, for example, those that are meant for single type of model in one stage of the AI lifecycle, a single organization, or a single method for explainability.
    \item Accessibility: We excluded papers not publicly available or tools only accessible upon special request. 
    \item Ambiguous roles: We excluded works using vague language such as ``AI practitioners'' or failing to identify the ``who'' and ``how'' of tool usage, if there was no clearly interpretable role or stage involved. 
    \item Roles external to organizational perspective: We excluded works that are not intended for leaders, designers, developers, deployers, end-users, or impacted communities (for example, works meant for governments or policy-makers). 
    % \Rachelmargincomment{Unsure if this is the best name}
\end{itemize} \\ 
         \bottomrule
    \end{tabular}
    \label{tab:inclusion_exclusion_criteria}
\end{table*}

\subsection{The AI Lifecycle and Stakeholders}

Following previous work~\cite{black2023toward, NIST2023, Ortega2024} we considered the following stages of designing, developing, and deploying an AI system (Table~\ref{tab:stage_table}).
%Here is the link to the visualization slide, I am still messing with it. https://docs.google.com/presentation/d/16CKZeEffnZj98NqwQAAyibxPk0nl6WctMuiiylzCRqI/edit?usp=sharing 
Furthermore, we borrowed from the AI actors included in the NIST Risk Management Framework (NIST AI RMF)~\cite{NIST2023} to consider the AI system stakeholder groups found in Table \ref{tab:role_table}.

% \begin{figure}[]
% \includegraphics[width=\linewidth]{figs/stage_fig.png}
% \centering
% \caption{Lifecycle Stage Definitions}
% \label{fig:stage_fig}
% \end{figure}

\begin{table*}[]
\centering
\scriptsize
\caption{Definitions for major stages in the AI Lifecycle}
\begin{tabular}
{cp{0.75\linewidth}}
\toprule
\textbf{Stage}
 & \textbf{Definition} \\ \midrule
\textbf{Value Proposition}           &  Value Proposition refers to a series of early investigations into the problem the AI system is designed to solve, what the business and policy requirements are, and whether including an AI component within the broader decision-making system leads to net-benefit over the status quo. \\ \midrule
\textbf{Problem Formulation}         &  Problem Formulation refers to the translation of business needs and requirements to technical choices, what types of methodologies will be used, and how the overall system will incorporate the AI system.   \\ \midrule
\textbf{Data Collection}        &  Data collection involves selecting, collecting, or compiling data to train and/or validate an AI model, and it involves making choices---or implicitly accepting previously made choices---about how to sample, label, and link data.  \\ \midrule
\textbf{Data Processing}         &  Data processing refers to the steps taken to make data usable by the ML model---for example, cleaning, standardizing, or normalizing data, including feature engineering. \\ \midrule
\textbf{Statistical Modeling}    &  Statistical Modeling consists of deciding what type of statistical model(s) to fit to the data, and how to perform the fitting. The latter requires choosing the learning rule and loss function, regularizers, hyperparameters' values, model selection methodology, and model selection metrics to be used. \\ \midrule
\textbf{Testing} & Testing refers to the process of verifying that the AI system functions correctly and produces outputs as expected. This stage involves ensuring that the code is implemented properly, free of critical bugs, and capable of handling various input scenarios without errors. Testing focuses on the technical robustness of the system, such as confirming that the model generates predictions, processes data as intended, and operates reliably in different environments.   \\ \midrule
\textbf{Validation }        &  Validation goes beyond testing to assess whether the system's performance meets specific criteria or thresholds necessary for deployment. This process evaluates the model's accuracy, fairness, bias mitigation, and alignment with ethical and regulatory standards. \\ \midrule
\textbf{Deployment}         &  Deployment refers to the process of deploying the model into a larger decision system.  \\ \midrule
\textbf{Monitoring}         &  Monitoring refers to how a model’s behavior is recorded and responded to over time to ensure there is no significant or unexpected degradation in performance and use over time.  \\ \bottomrule
\end{tabular}
\label{tab:stage_table}
\end{table*}

\begin{table*}[]
\centering
\caption{Role definitions for major stakeholders in the AI lifecycle}
\scriptsize
\begin{tabular}
{cp{0.75\linewidth}}
\toprule
\textbf{Role}
 & \textbf{Definition} \\ \midrule
\textbf{Leaders}           &  This role makes strategic decisions about the goals, objectives, and missions of the organization and the high-level policies to implement those goals. Typically, leaders are at the top of the organizational hierarchy or structure and include roles such as C-suite executives, and leaders of units or teams who are delegated specific responsibilities. In addition to C-suite executives, these roles can include senior management, unit or group leaders, policy leaders, or team leaders.  \\ \midrule
\textbf{Designers}         &  This role oversees the reframing of business and policy goals into a technical problem, aligning these problems with stakeholder needs in a particular context of use, and envisioning the user experience. Example roles include domain experts, AI designers, product managers, compliance experts, human factors experts, UX designers, and others who are familiar with doing cross-functional work.  \\ \midrule
\textbf{Developers}        &  This role uses programming or other technical knowledge to prepare, curate, and engineer data and to train AI models to perform the specified task. Example roles include data engineers, modelers, model engineers, data scientists, AI developers, software engineers, and systems engineers.  \\ \midrule
\textbf{Deployers}         &  The deployer role verifies and validates the model beyond the training and test data, pilots the model, checks compatibility with existing systems, and collaborates with designers to evaluate the user experience. Roles include testing and evaluation experts, auditors, impact assessors, and system integrators.   \\ \midrule
\textbf{End-users}         &  This role represents any individual who utilize the output of the AI model to contribute to an organizational goal (for example, making decisions or automating workflows and practices).  \\ \midrule
\textbf{Impacted Communities} & This stakeholder group represents any individual or community impacted directly or indirectly by the AI system’s operation and their advocates and representatives. Roles include groups that may be harmed by the model, advocacy groups, and the broader public.    \\ \bottomrule
\end{tabular}
\label{tab:role_table}
\end{table*}

\subsection{Exploratory Data Analysis}
After applying the inclusion and exclusion criteria, we began to assess the final set of papers according to the roles and lifecycle stages they addressed. 
Note that we chose to focus on the lifecycle stages that are \emph{addressed} rather than the stages where the tool is \emph{used}; for instance, although \rkedit{Model Cards}~\cite{Mitchell2019} are relevant to end-users after deployment, they address ``statistical modeling.''
Taking the stakeholder groups as rows and the stages as columns, we constructed a matrix, organizing the RAI tools into the rows and columns they address.
Our goal was to understand the relative abundance or lack of lifecycle stages and roles addressed in responsible AI governance. 
% To that end, we began categorizing the papers by developing a table of the roles and lifecycles stages that were addressed by each paper.

We were also interested in exploring whether the tools introduced in the literature were validated in any way for usability or effectiveness before publishing. Specifically, we evaluated tools and processes to determine whether they included hypothetical case studies,
experimental results, pilot programs, or even reports of results in real-world use.

To ensure consistency in labeling, two of the co-authors independently labeled a small subset of the dataset, focusing on lifecycle stages, roles, and validation criteria. Overall, this resulted in a Krippendorff's alpha score of .883 for stage labels and .824 for role labels, indicating substantial agreement. With these values, we were confident in our alignment and we moved forward with labeling the rest of the data individually, maintaining frequent check-ins to address any uncertainties.

\subsection{Limitations}
% Possible biases in literature collection.\\
% Constraints in defining roles or lifecycle stages.
Despite its contributions, the methodology used in this literature review has some limitations. First, potential biases in the literature collection may have influenced our findings. Our categorization of papers and tools relies on the availability and accessibility of published work, which may overrepresent well-funded research areas or regions. \rkedit{Second}, our definitions of lifecycle stages and roles, while informed by existing literature
%, are necessarily constrained by our framework’s scope and 
may not capture the full complexity of real-world AI development practices.
\rkedit{Third}, the boundaries between stages and roles can be fluid, complicating efforts to assign responsibilities and tools to specific cells in our matrix. For example, developers often contribute to problem formulation and validation, even if these are not traditionally considered part of their primary responsibilities. Similarly, stages such as deployment and monitoring often overlap, particularly in iterative development processes.
\rkedit{Fourth, by using the number of tools available for each stage and stakeholder group, we do not consider the popularity of tools in real-world practices. Towards, this end,}
\bkedit{we also considered citation counts as a proxy for academic visibility and impact (\rkedit{see Appendix~\ref{app:citation_count}} Figure~\ref{fig:citation_count}), but this measure does not fully capture the practical use or popularity of tools in real-world applications. 
For some tools, we may be able to look at additional proxy metrics, such as GitHub forks. However, in addition to such metrics being imperfect proxies for impact, they are not applicable across all the tools in our literature review. \rkedit{For these reasons}, we chose not to assess the popularity of the tools reported, but rather focused our corpus collection using publicly available information on the Web. }
\section{Findings}\label{sec:findings}

We first analyze the AI lifecycle stages and stakeholders addressed by the RAI tools in our systematic literature review.
Then, we explore the intersection between stakeholders and stages, quantifying how well each intersection is covered in the literature both when considering all tools and when filtering by tools that have been validated.
Finally, we explore whether and how existing tools are validated.
% Our findings highlight roles and lifecycle stages that are overrepresented and, conversely, areas that are underrepresented and may benefit from further research.
% Furthermore, we showcase the need for a more comprehensive validation of the efficacy and reliability of frameworks and tools.
% Additionally, we explored the existing relations between roles and lifecycle stages. 

\subsection{Lifecycle Stages} \label{sec:lifecycle_stages}
\begin{figure*}[]
     \centering
     \includegraphics[width=0.875\linewidth]{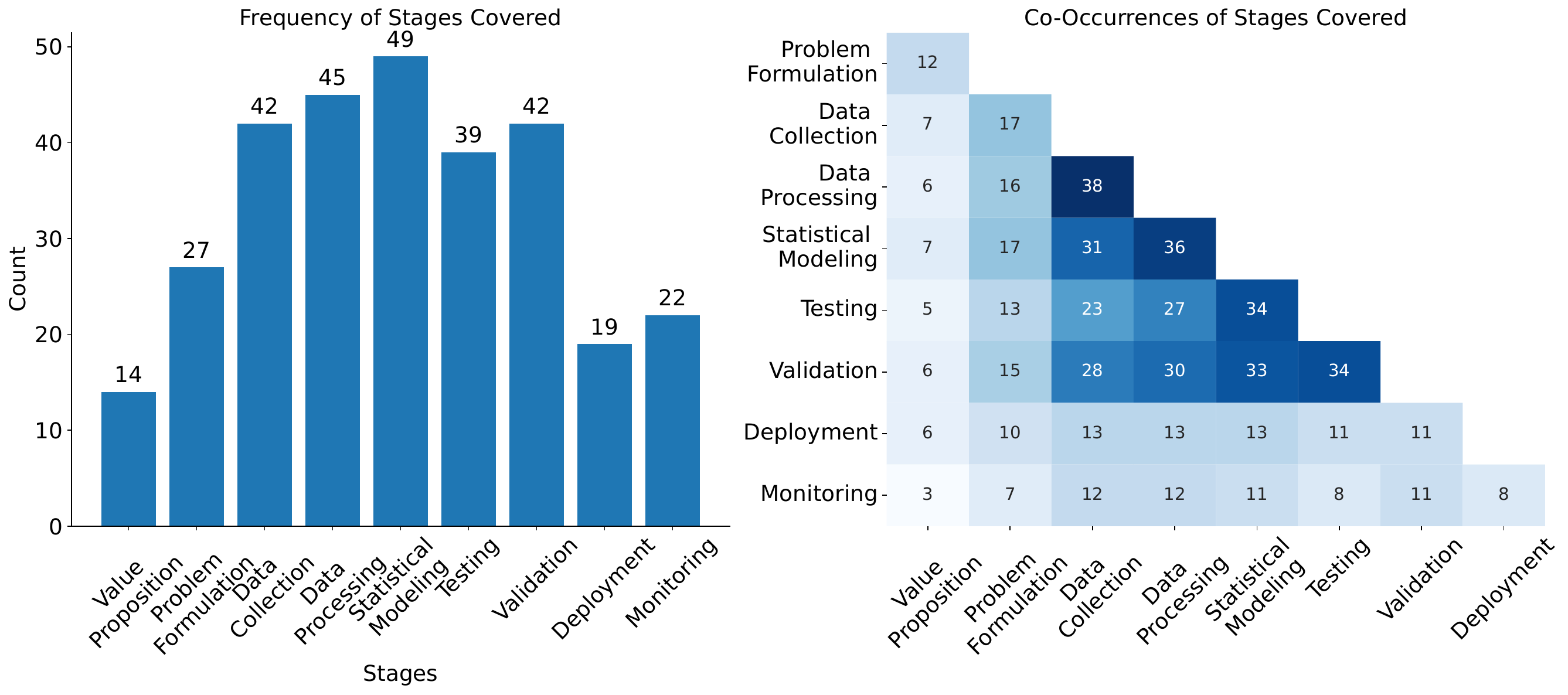}
     \caption{The distribution of stages present for validated tools (left) and the co-occurrence of pairs of stages present for validated tools (right).}
     \label{fig:stages_sbs}
\end{figure*}

Figure~\ref{fig:stages_sbs} (left) contains the lifecycle stages addressed by the tools from our systematic literature review.
The most frequently represented lifecycle stages include ``Data Collection'' (50.0\% of all validated tools), ``Data Processing'' (53.7\% of all validated tools), ``Statistical Modeling'' (59.8\% of all validated tools), and ``Testing'' (47.6\% of all validated tools), and ``Validation'' and (51.2\% of all validated tools). 
Tools for these stages often target tasks such as fairness auditing~\citep[e.g.,][]{Weerts2023, Bellamy2019, Saleiro2018}, explainable AI~\citep[e.g.,][]{Nori2019, h2o2024, Tenney2020}, and benchmarking~\citep[e.g.,][]{Rauber2017}.
This overrepresentation can be attributed to the technical nature of these tasks and the strong emphasis on technical tools in responsible AI research, a trend also found previously by Ojewale et al.~\cite{Ojewale2024} and Black et al.~\cite{black2023toward}. 
% In addition, we hypothesize that there is a general perception that these stages are critical to achieving measurable outcomes such as fairness, transparency, and precision, which constitute some of the key ethical values of RAI. 

Conversely, 
% ``Problem Formulation'', ``Deployment'', and ``Monitoring'' were all less well represented with 35.4\%, 26.8\%, and 29.3\% respectively.
``Value Proposition'' and ``Problem Formulation'' were much less represented, covering only 17.1\% and 32.9\% of all validated tools, respectively.
The comparative neglect of these early AI lifecycle stages in our systematic literature review suggests a broader systemic issue discussed by many RAI researchers and practitioners~\cite{kawakami2024situate, coston2023validity, passi2019problem, raji2022fallacy}: many RAI challenges originate from the early stages of AI development, but RAI is often retrofitted rather than embedded from the start. 
% We hypothesize that the promises of AI---increasing economic efficiency, boosting productivity, and allowing for human capital reinvestment---are both implicitly assumed and perceived to be enough to provide a ``green light'' for a project to move through subsequent stages. 
Ethics concerns are often seen as something that can and should be addressed later.

Additionally, ``Deployment'' and ``Monitoring'' are underrepresented, with 23.2\%, and 26.8\% of all validated tools addressing these stages. 
This gap may point to a broader issue within the RAI landscape: the lack of tools that address the ongoing lifecycle needs of AI systems after their deployment. 
% Unlike data collection or modeling, deployment and monitoring involve continuous oversight and adaptation, yet they remain significantly underserved by the current tool ecosystem.
As a result, many present-day deployment and monitoring practices are not standardized, instead relying on the perspectives and expertise of the actors involved in these stages~\cite{cheng2022heterogeneity, eslami2017careful}. 
However, this can result in heterogeneous decision-making---for example, senior call-workers are more likely to screen in referrals in the context of child abuse hotline screening~\cite{cheng2022heterogeneity}.
This variance can have downstream implications, especially in high-stakes domains.

% \begin{figure}[H]
% \centering
% \begin{minipage}{.4\textwidth}
%   \centering
%   \includegraphics[width=\linewidth]{figs/stages.png}
%   \captionof{figure}{Distribution of Stages}
%   \label{fig:stages}
% \end{minipage}%
% \begin{minipage}{.4\textwidth}
%   \centering
%   \includegraphics[width=\linewidth]{figs/stages_stages.png}
%   \captionof{figure}{Co-occurrence of Stages}
%   \label{fig:stages_stages}
% \end{minipage}
% \end{figure}

Finally, we count how many RAI tools address each possible pair of stages (Figure~\ref{fig:stages_sbs}, right). 
We observe that while there are many RAI tools that cover pairs of stages in the middle of the AI lifecycle, there are comparatively less processes that cover the AI lifecycle from end-to-end. Notably, there are only three validated tools that cover both value proposition and monitoring. Of these three tools, two cover \emph{all} of the AI lifecycle stages in Table~\ref{tab:stage_table}.
The first tool is the Government Accountability Office (GAO's) Accountability Framework~\cite{GAO2021}, \rkedit{which contains} key questions and processes for auditors.
The second tool is Microsoft's AI Fairness Checklist~\cite{Madaio2020} \rkedit{which contains} questions for organizations to consider to ensure that their AI systems are fair.

\subsection{Stakeholders} \label{sec:stakeholders}
\begin{figure*}[]
     \centering
     \includegraphics[width=0.875\linewidth]{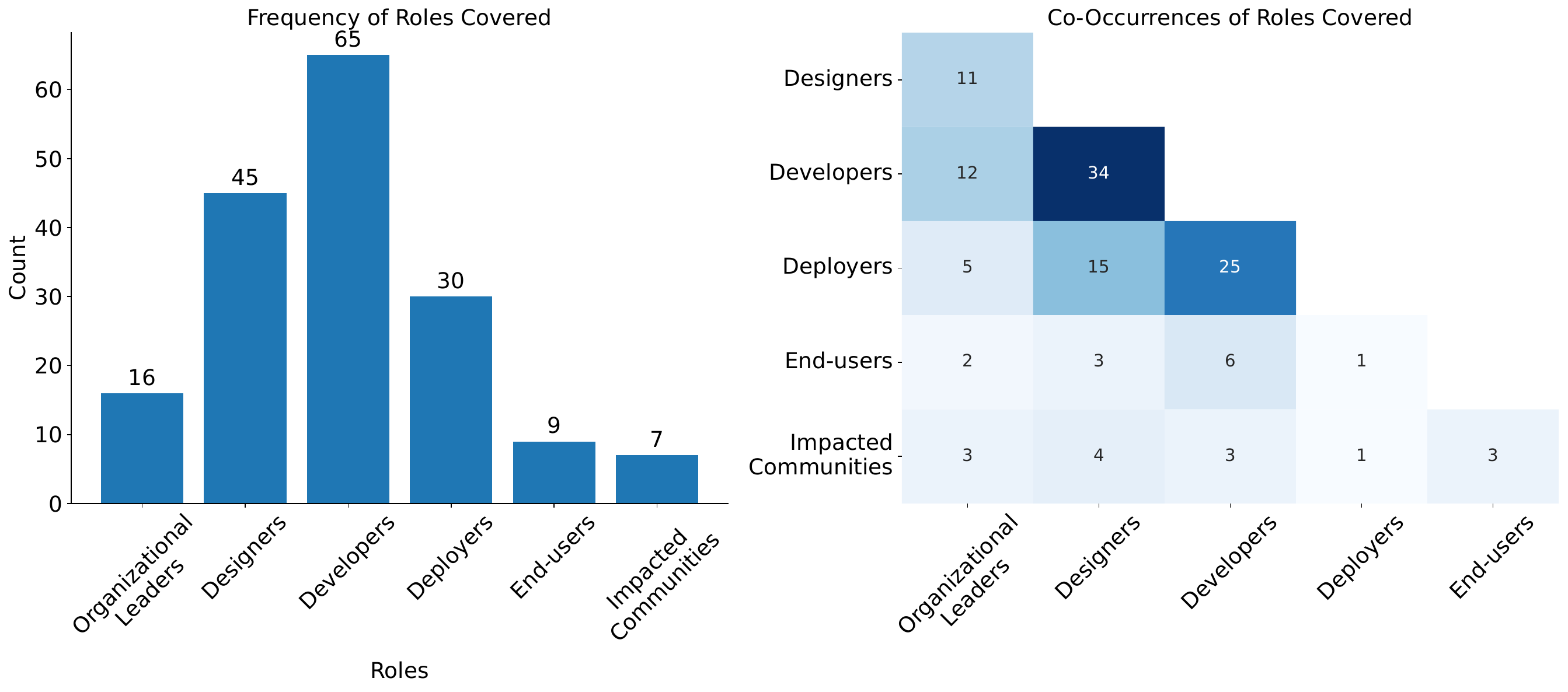}
     \caption{The distribution of roles present for validated tools (left), and the co-occurrence of pairs of roles present for validated tools (right).}
     \label{fig:roles_sbs}
\end{figure*}

Our analysis uncovered that some stakeholders and stages within AI governance were overrepresented and others were underexplored in the literature. We highlight the results in Figure~\ref{fig:roles_sbs} (left).
Overall, ``Developers'' were the most frequently addressed role, with 78.0\% of all validated tools addressing aspects of their responsibilities. ``Designers'' were also well represented, with 53.7\% of all validated artifacts highlighting tools and processes for designers. 
We suspect that this overrepresentation is a reflection of the fact that these stakeholder groups are often perceived to be the main actors in AI development. 
As the main technical stakeholders, it is unsurprising that they most often coexist with the technical stages discussed above. 
These pairings of, and the emphasis on, technical roles and stages could be leading to the stakeholder-stage overrepresentation that we have identified. 

Conversely, ``Organizational Leaders'' were relatively underrepresented with only addressed with 18.3\% of all validated artifacts providing tools and processes for their roles. 
Organizational leaders are a powerful stakeholder group who have the ability to set overarching goals, make strategic decisions, and establish the working culture.
The high-level choices made by organizational leaders are perpetuated down to the stakeholder groups that interact more directly with AI systems, such as designers, developers, and deployers.
The lack of RAI tools for organizational leaders suggests that they are choosing to use AI within their organizations without fully understanding its benefits or risks. 
% The detailed steps required to ensure good AI governance then becomes the responsibility of the designers and developers.
On the other hand, ``End-users'' and ``Impacted Communities'' were vastly underrepresented when compared to the top two roles, with only 11.0\% and 9.8\% of all validated tools addressing those roles, respectively.
This statistic shows that similar to the findings by Kawakami et al.~\cite{kawakami2024responsible}, the landscape of RAI tools currently treats end-users and impacted communities as an afterthought.

Additionally, we count the number of RAI tools that are designed for each possible pair of stakeholders present in Table~\ref{tab:role_table}. The co-occurrences are shown in Figure~\ref{fig:roles_sbs} (right).
We find that there are many tools that address the combination of ``Designer'' and ``Developer'', and ``Developer'' and ``Deployer''. 
There are also a moderate number of tools that address the combination of ``Organizational Leader'' and ``Designer'', ``Organizational Leader'' and ``Developer'', and ``Designer'' and ``Deployer''.
However, there is a lack of tools that are designed for ``End-users'' and ``Impacted Communities'' with other stakeholder groups.
One prominent validated tool that addresses leaders, end-users, and impacted communities is Model Cards~\cite{Mitchell2019}.

% \Rachelmargincomment{What is the main takeaway from the co-occurrence matrix? If it isn't that interesting, maybe we should remove it.}

% \begin{figure}[H]
% \centering
% \begin{minipage}{.4\textwidth}
%   \centering
%   \includegraphics[width=\linewidth]{figs/roles.png}
%   \captionof{figure}{Distribution of Roles}
%   \label{fig:roles}
% \end{minipage}%
% \begin{minipage}{.4\textwidth}
%   \centering
%   \includegraphics[width=\linewidth]{figs/roles_roles.png}
%   \captionof{figure}{Co-occurrence of Roles}
%   \label{fig:roles_roles}
% \end{minipage}
% \end{figure}

\subsection{Stakeholders and Stages}
To better understand the interactions between roles and lifecycle stages, we analyzed their co-occurrences within the dataset.
Figure~\ref{fig:roles_stages_sbs} (right) represents all the tools from our systematic literature review, while Figure~\ref{fig:roles_stages_sbs} (left) represents only the tools that have been validated in some form.
% The heatmaps reveal key patterns in pairings and also in the lack of pairings. 
We see that the most frequent pairings involved developers with ``Data Collection'' and ``Statistical Modeling'', as well as designers with ``Problem Formulation''. These findings align with the roles and stages traditionally emphasized in technical aspects of AI development. 

Conversely, intersections involving ``Impacted Communities'' and any of the lifecycle stages were notably sparse. 
A similar pattern exists for ``End-users'' at all stages, albeit at a slightly higher frequency. 
In particular, there are little tools available to these stakeholder groups in the earlier stages of the AI lifecycle.
Perhaps most notably, there are \emph{no} validated tools that encompass both end-users and value proposition, and \emph{one} validated tool (Community Jury~\cite{MSCJ2022}) that encompasses both impacted communities and value proposition (Figure~\ref{fig:roles_stages_sbs}, left).
Current RAI tools do not empower end-users and impacted communities to voice whether an AI system would be useful for them (in the case of end-users) want an algorithm to affect a certain aspect of their life (in the case of impacted communities).
This finding highlights that the value of an AI system is primarily perceived through the lens of higher-level organizational goals rather than individual experiences, end-users' and impacted communities' have a lack of power.
As Kawakami et al.~\cite{kawakami2024responsible} also argue, many organizations developing AI systems treat end-users and impacted communities as an afterthought.
% https://ojs.aaai.org/index.php/AIES/article/view/31669/33836
% - why is this happening. end-users and impacted communities are an afterthought for an organization developing AI systems. focus on higher-level, organizational level goals rather than the individual experience with an algorithm.
% - 
% This finding suggests that current RAI tools do not empower end-users and impacted communities to voice whether an AI system would be useful for them (in the case of end-users) want an algorithm to affect a certain aspect of their life.
% The organizational structures around AI systems further reinforce pre-existing inequalities: end-users and impacted communities, who are often the stakeholder groups with the least power, are given no say in whether an algorithm is valuable to them.

The previous claim is further supported by the fact that of the small number of RAI tools available to end-users and impacted communities, many are focused on monitoring, especially for impacted communities. (Figure~\ref{fig:roles_stages_sbs}, left).
This means that current tools only help end-users and impacted communities take action towards RAI \emph{after} a model has been deployed. 
However, after a AI system is deployed, undesired effects on end-users and impacted communities may have already begun to take place.
To illustrate, two tools for end-users and impacted communities in the monitoring stage include a black-box method to detect whether a certain user's texts were used to train a model~\cite{Song2019} and a browser extension that informs people about the entities that might be targeting political ads towards them~\cite{WTM2017}. 
If end-users and impacted communities discover through these tools that their information has been unwillingly used in model training or that they have been unknowingly targeted by a political entity, the harm has already occurred: the individual's privacy has already been violated or the individual has not been provided the transparency they deserve.
End-users and impacted communities often the least powerful stakeholder groups within the AI lifecycle~\cite{bondi2021envisioning}, which can result in them being given a voice only after harm has already occurred, rather than actively participating in efforts to prevent harm from the outset.
% This suggests that while technical and managerial roles are well-covered, tools that engage external stakeholders or consider the long-term societal impact of AI systems are 
% % less incentivized and therefore 
% lacking. 

% \begin{figure}[H]
%     \centering
%     \includegraphics[width=1\linewidth]{figs/roles_stages.png}
%     \caption{Co-occurrence of roles and stages, filtered by tools that have been validated}
%     \label{fig:roles_stages}
% \end{figure}

\subsection{Tool Table}
In Appendix~\ref{app:tool_table} Table~\ref{tab:validated_tools}, we have constructed a tool table that includes the available, validated tools and processes for each lifecycle stage and stakeholder intersection. Additionally, we have provided Table~\ref{tab:ex_tools} of example tools for each stakeholder-stage intersection.

% \begin{landscape}
\begin{table*}[]
\centering
% \renewcommand{\arraystretch}{1.5} 
% \setlength{\tabcolsep}{5pt}  
% \footnotesize
% \scriptsize
\tiny
\caption{Examples of validated tools for each lifecycle stage and stakeholder}
\begin{tabular}
{p{0.075\linewidth}p{0.075\linewidth}p{0.075\linewidth}p{0.075\linewidth}p{0.075\linewidth}p{0.075\linewidth}p{0.075\linewidth}p{0.075\linewidth}p{0.075\linewidth}p{0.075\linewidth}}
\toprule
 & \textbf{Value Proposition}
 & \textbf{Problem Formulation}
 & \textbf{Data Collection}
 & \textbf{Data Processing}
 & \textbf{Statistical Modeling}
 & \textbf{Testing}
 & \textbf{Validation}
 & \textbf{Deployment} 
 & \textbf{Monitoring}\\ \midrule

\textbf{Leaders} 
& \raggedright An Approach to Organizational Deployment of Inscrutable Artificial Intelligence Systems \cite{Asatiani2021} 
\newline \newline \emph{Provides a framework for assessing organizational readiness and capability to deploy AI responsibly} 
& \raggedright  A Lifecycle Approach for Artificial Intelligence Ethics in Energy Systems \cite{El-Haber2024} 
\newline \newline \emph{Outlines a comprehensive approach for incorporating ethical considerations into AI projects}
& \raggedright Guidance on the Assurance of Machine Learning in Autonomous Systems \cite{Hawkins2021} 
\newline \newline \emph{Offers specific guidance for certifying ML systems in high-stakes applications}
& \raggedright Model Cards \cite{Mitchell2019}  
\newline \newline \emph{Presents a template for summarizing key details of AI models, including intended use and potential risks}
& \raggedright Aequitas \cite{Saleiro2018} 
\newline \newline \emph{Evaluates fairness across groups and ensures equitable outcomes using metrics and dashboards}
&  \raggedright Assured AI reference architecture \cite{Tyler2024} 
\newline \newline \emph{Defines a reference architecture to ensure AI systems are trustworthy, explainable, and aligned with organizational goals}
& \raggedright From ethical AI frameworks to tools \cite{Prem2023} 
\newline \newline \emph{Maps high-level ethical AI principles to actionable tools, enabling practical implementation of ethical AI} 
& \raggedright Corporate digital responsibility \cite{Lobschat2021} 
\newline \newline \emph{Aligns AI deployment with corporate digital responsibility initiatives to support sustainable and ethical practices}
& A Framework for Evaluating and Disclosing the ESG Related Impacts of AI with the SDGs \cite{Saetra2021} 
\newline \newline \emph{Evaluates AI’s environmental, social, and governance (ESG) impacts through alignment with Sustainable Development Goals }\\ \midrule

\textbf{Designers} 
& \raggedright ALTAI \cite{EUAI2020} 
\newline \newline \emph{Guides designers through ethical self-assessment and understanding AI's societal impact via interactive tools}
& \raggedright Prescriptive and Descriptive Approaches to Machine-Learning Transparency \cite{Adkins2022} 
\newline \newline \emph{Provides transparency guidelines and case studies to help designers articulate and address AI opacity}
& \raggedright Decision Provenance \cite{Singh2019}  
\newline \newline \emph{Tracks and explains the origins of decisions in ML pipelines, ensuring transparency for stakeholders}
&  \raggedright AI Fairness 360 \cite{Bellamy2019} 
\newline \newline \emph{TOffers algorithms and tools to reduce bias in data preprocessing and model training}
& \raggedright Hybrid human-machine analyses for characterizing system failure \cite{Nushi2018} 
\newline \newline \emph{Combines human oversight with ML results to detect failure points and refine system performance}
& \raggedright VerifAI \cite{Dreossi2019} 
\newline \newline \emph{Simulates diverse real-world scenarios to evaluate robustness and safety of AI models}
& \raggedright Behavioral Use Licensing for Responsible AI \cite{Contractor2022} 
\newline \newline \emph{Introduces licensing frameworks that enforce ethical AI behaviors in deployment}
& \raggedright Model AI Governance Framework \cite{PDPC2020} 
\newline \newline \emph{Provides actionable steps for organizations to build robust AI governance practices}
& Artificial Intelligence Ethics and Safety \cite{Corra2021} 
\newline \newline \emph{Suggests ethical principles for AI projects in the design phase, focusing on community benefit}\\ \midrule

\textbf{Developers} 
& \raggedright Co-Designing Checklists \cite{Madaio2020} 
\newline \newline \emph{Facilitates co-creation of development checklists to address fairness, bias, and accountability issues}
& \raggedright FactSheets \cite{Arnold2019} 
\newline \newline \emph{Summarizes AI system specifications, including intended use, limitations, and safety measures}
& \raggedright Datasheets for datasets \cite{Gebru2021} 
\newline \newline \emph{Documents datasets with metadata on collection methods, ethical considerations, and intended uses}
& \raggedright Conscientious Classification \cite{dAlessandro2017} 
\newline \newline \emph{Proposes techniques for classifying data with fairness constraints, balancing accuracy and equity}
& \raggedright FairLens \cite{Panigutti2021} 
\newline \newline \emph{Identifies sources of bias in datasets, aiding in debiasing efforts during feature engineering}
& \raggedright LiFT \cite{Vasudevan2020} 
\newline \newline \emph{Tests fairness and robustness of ML models using automated workflows and visualization tools}
& \raggedright Fairness-enhancing Interventions \cite{Friedler2019} 
\newline \newline \emph{Provides fairness-enhancing interventions, including pre- and post-processing methods}
& \raggedright IBM Watson Studio \cite{IBM2024} 
\newline \newline \emph{Enables responsible development workflows through prebuilt tools and assessment metrics}
& The medical algorithmic audit \cite{Liu2022} 
\newline \newline \emph{Audits algorithmic outputs, identifying disparities in decision-making}\\ \midrule

\textbf{Deployers} 
& \raggedright Corporate digital responsibility \cite{Lobschat2021} 
\newline \newline \emph{Aligns deployment strategies with organizational ethics and corporate responsibility goals}
& \raggedright TensorFlow Responsible AI \cite{TF2025} 
\newline \newline \emph{Offers Responsible AI tools integrated with TensorFlow for bias mitigation and transparency}
& \raggedright Decentralized Big Data Auditing \cite{Yu2019} 
\newline \newline \emph{Implements decentralized auditing processes for ensuring the security and accuracy of data pipelines}
& \raggedright FairPy \cite{Viswanath2023} 
\newline \newline \emph{Provides preprocessing algorithms to identify and reduce potential biases in datasets}
& \raggedright Fairtest \cite{Tramer2017} 
\newline \newline \emph{Tests ML models for fairness using specific tools that simulate diverse user interactions}
& \raggedright The Language Interpretability Tool \cite{Tenney2020} 
\newline \newline \emph{Visualizes model interpretability, offering insights into decision boundaries and errors}
& \raggedright From ethical AI frameworks to tools \cite{Prem2023} 
\newline \newline  \emph{Links organizational ethical frameworks to specific tools, aiding in post-deployment validation}
& \raggedright PyRIT \cite{Munoz2024} 
\newline \newline \emph{Tracks AI practices and analyzes responsible development actions across lifecycle stages}
& FairTest \cite{Tramer2015} 
\newline \newline \emph{Identifies critical algorithmic errors during deployment, focusing on reducing harm}\\ 
\midrule

\textbf{End-users}     
% & \raggedright \cite{}  
& \raggedright []  
& \raggedright Model Cards \cite{Mitchell2019}  
\newline \newline \emph{Simplifies understanding of AI models by providing key details on risks, performance, and scope} 
& \raggedright Decision Provenance \cite{Singh2019} 
\newline \newline \emph{Offers explanations of AI-driven decisions, increasing trust and usability for end-users}  
& \raggedright Towards Intersectional Feminist and Participatory ML \cite{Suresh2022} 
\newline \newline \emph{Promotes participatory development, incorporating diverse perspectives into ML design}  
& \raggedright A maturity assessment framework for conversational AI development platforms \cite{Aronsson2021} 
\newline \newline  \emph{Evaluates usability of conversational AI systems to improve accessibility and user experience }
& \raggedright Hybrid human-machine analyses for characterizing system failure \cite{Nushi2018} 
\newline \newline \emph{Combines human judgment and machine analysis to understand and address system errors}
& \raggedright Model Cards \cite{Mitchell2019} 
\newline \newline \emph{Summarizes AI models in ways that are interpretable and actionable for end-users} 
& \raggedright A Lifecycle Approach for Artificial Intelligence Ethics \cite{El-Haber2024}  
\newline \newline \emph{Guides ethical recommendations for AI usage through a lifecycle framework, with an emphasis on end-user systems}
& Discovering and Validating AI Errors With Crowdsourced Failure Reports \cite{Cabrera2021}  
\newline \newline \emph{Crowdsources user feedback to identify and validate AI failures during post-deployment} \\ 
\midrule

\textbf{Impacted Communities}  
& \raggedright Microsoft Community jury \cite{MSCJ2022} 
\newline \newline \emph{Incorporates community voices into AI decisions through jury-based participatory approaches} 
& \raggedright Practical fundamental rights impact assessments \cite{Janssen2022}  
\newline \newline \emph{Provides frameworks for assessing fundamental rights impacts of AI systems on vulnerable groups}
& \raggedright A Lifecycle Approach for Artificial Intelligence Ethics \cite{El-Haber2024} 
\newline \newline \emph{Ensures AI projects align with societal values and promote positive outcomes for communities} 
& \raggedright Model Cards \cite{Mitchell2019} 
\newline \newline \emph{Communicates essential model details, such as intended use and limitations, to the public } 
& \raggedright UnBias Fairness Toolkit \cite{Lane2018} 
\newline \newline \emph{Engages communities in understanding fairness and addressing biases in algorithms}
& \raggedright Testing Concerns about Technology’s Behavioral Impacts \cite{Matias2022}  
\newline \newline \emph{Examines potential community harms of technology and offers mitigation strategies}
& \raggedright Model Cards \cite{Mitchell2019} 
\newline \newline \emph{Documents AI systems accessibly, ensuring impacted groups can understand and question use cases}
& \raggedright A Lifecycle Approach for Artificial Intelligence Ethics \cite{El-Haber2024} 
\newline \newline  \emph{Provides a structure to enable discussion for ensuring AI systems align with ethical standards and societal values.}
& Auditing Data Provenance in Text-Generation Models \cite{Song2019} 
\newline \newline \emph{Audits data provenance in generative text systems to evaluate risks of misinformation}\\ \bottomrule
\end{tabular}
\label{tab:ex_tools}
\end{table*}
% \end{landscape}

\subsection{Tool Validation} \label{sec:tool_validation}
Our analysis found that a significant portion of the tools and practices identified in the literature lack validation of any kind (\rkedit{142} of \rkedit{224}, \rkedit{63.4}\%) and only a small portion presented evidence of validation (82 of \rkedit{224}, \rkedit{36.6}\%). 
Among those that were validated, validation methods include a hypothetical case study~\citep[e.g.][]{dAlessandro2017, Moon2019, Vidgen2019}, testing code~\citep[e.g.][]{Tyler2024, Ryffel2018, Friedler2019}, or a real-world pilot~\citep[e.g.][]{IBM2024, Ballard2019, Gebru2021}.
Some of the unvalidated RAI tools are built by RAI experts~\citep[e.g.][]{ODI2021, CDT2017, DC2021}, include consultations and collaborations with relevant people and groups ~\citep[e.g.][]{ACLU2020, ECP2019, UNICEF2024}, and provide mechanisms for tool-users to give feedback~\citep[e.g.][]{DD2022, GovEx2018, Doteveryone2019}.
Consequently, builders of such tools may not feel the need to undergo a validation process given the rigorous development process and existing monitoring strategies.

% However, we found that many tools did not provide sufficient evidence of practical application, indicating a need for more rigorous evaluation methodologies in this domain.
% Of the RAI tools in our systematic literature review, only 35\% of the tools contain any evidence that the tool creators checked the tool's validity, applicability, or usability.
% Examples of validation include a hypothetical case study~\citep[e.g.][]{dAlessandro2017, Moon2019, Vidgen2019}, testing code~\citep[e.g.][]{Tyler2024, Ryffel2018, Friedler2019}, or a real-world pilot~\citep[e.g.][]{IBM2024, Ballard2019, Gebru2021}.
% Some of the unvalidated RAI tools are built by RAI experts~\citep[e.g.][]{ODI2021, CDT2017, DC2021}, include consultations and collaborations with relevant people and groups ~\citep[e.g.][]{ACLU2020, ECP2019, UNICEF2024}, and provide mechanisms for tool-users to give feedback~\citep[e.g.][]{DD2022, GovEx2018, Doteveryone2019}.
% Builders of such tools may not feel the need to undergo a validation process.

\section{Discussion}\label{sec:discussion}
% Practical guidance for researchers and practitioners to address gaps.\\
% Importance of interdisciplinary efforts to fill underrepresented areas.

% Despite organizations increasing eagerness to adopt AI---the percentage of organizations surveyed by McKinsey adopting AI jumped from 55\% in 2023 to 72\%~\cite{}---leaders are still largely AI illiterate.
% % https://www.mckinsey.com/capabilities/quantumblack/our-insights/the-state-of-ai
% Notably, previous research shows low AI literacy among executives, with only 3\% of executives at S\&P 500 companies viewed as AI literate~\cite{}
% % https://sloanreview.mit.edu/article/why-executives-cant-get-comfortable-with-ai/

There are several implications for RAI development that can be drawn from our findings. They center around improper use of RAI tools and fragmented approaches to RAI. We elaborate on these below. 
% \TODO{write a summary of the discussion}

% Organizational leaders may struggle to enforce ethical standards because of their low AI literacy~\cite{} and the lack of actionable tools to operationalize governance. In turn, the overall accountability of AI systems is undermined. 
% This gap can result in a disconnect between high-level principles and their practical implementation, leaving ethical considerations as aspirational rather than actionable. Leaders may find it challenging to translate organizational values into concrete practices, particularly in complex, fast-moving development cycles. This lack of operational clarity can create inconsistencies in decision-making and accountability, increasing the risk of ethical oversights. Furthermore, without tools to monitor and measure the impact of their governance strategies, leaders are left unable to assess or refine their approaches, perpetuating a cycle of ineffective oversight that compromises trust and responsibility within AI-driven organizations. 

\xhdr{Improper use of RAI tools}
% The lack of validated processes for AI governance means that organizations searching for RAI artifacts to use cannot be sure of the usability, applicability, or validity of what is available. 
% If an organization decides to employ an unvalidated process, they may encounter issues with the usability of the process, finding it difficult to apply the process to their particular AI use case.
% Even worse, an organization employing a process that is inapplicable or invalid for their use case can lead to a false sense of confidence in adhering to ethical values.
% To illustrate, previous work has shown that developers can over-rely on explainability and interpretability tools~\cite{kumar2021shapley, kaur2020interpreting}.
Many RAI tools are not validated, which can result in publicly available artifacts that are unusable, prone to misuse, or ineffective, consequently providing a false sense of assurance surrounding AI systems.
% In Section~\ref{sec:tool_validation}, we reported that many RAI tools are not validated.
By a tool ``not being validated'', we mean that validation is not mentioned \emph{within the tool's documentation}, although researchers and third parties may have done validation studies later~\citep[e.g.,][]{deng2022exploring, kaur2020interpreting, nunes2022using}.
However, given the already-existing barriers to organizations adopting RAI practices, such as a fast-paced work environment~\cite{berman2024scoping, kaur2020interpreting, Tenney2020}, we hypothesize that organizations are unlikely to search for additional evidence of RAI tool validation if it is not already mentioned within the tool itself. 
Consequently, an organization considering an unvalidated RAI tool likely has no information regarding a tool’s effectiveness, that is, whether using the RAI tool delivers the claimed benefit in practice~\cite{cartwright2009thing}.
% Finding --> many tools are not validated
% Show no evidence of validation in the documentation of the tool (although researchers may have done validation studies later)
% This means that an organization looking at a tool has no idea about a tool’s effectiveness → does the tool achieve what it says it will achieve 

Attempting to use unvalidated RAI tools poses a myriad of issues.
First, an unvalidated RAI tool can be difficult to use~\cite{lee2021landscape}.
Furthermore, it is hard to pinpoint \emph{why} there are difficulties with using the tool. For example, the difficulty may be rooted in the tool's design, but could also be the result of a poor onboarding process and a lack of guidance for practitioners~\cite{berman2024scoping}.
Conversely, while an unvalidated RAI tool may be highly usable, without comprehensive checks on how the tool should and should not be used, practitioners can easily become prone to \emph{over-relying} on it. For example, previous work has shown that developers can over-rely on explainability and interpretability tools~\cite{kumar2021shapley, kaur2020interpreting}.
Additionally, an unvalidated RAI tool may not be suitable for existing organizational practices.
There are already many barriers to RAI adoption including resource constraints~\cite{berman2024scoping} and organizational culture~\cite{rakova2021responsible, varanasi2023currently}.
If a tool causes too much friction with already-present workflows, it may be more difficult to incentivize stakeholders to adopt the tool.
% This poses a myriad of issues, for instance
% Are tools easy to use by the stakeholders?
% On one hand, a tool might not be usable
% Difficulty of pinpointing where the source of unusability (?) is
% Design flaw of the tool?
% Onboarding / lack of guidance on how to use the tool?
% On the other hand, it might be very usable, but consequently cause practitioners to misuse it
% Ex. overreliance on interpretability tools
% Are tools suitable for existing organizational practices?
% Already there are barriers to adopting RAI (CITE)
% If tools are not designed for organizational workflows, it might be harder to incentivize people to adopt them

Even when we consider validated RAI tools, we find that many of the validations are removed from real-world contexts.
For example, many RAI tools are validated through a hypothetical case study~\citep[e.g.][]{dAlessandro2017, Moon2019, Vidgen2019} that illustrates how the tool should be used.
However, such forms of validation leave the critical question---what role can the RAI tool can play in the real-world?---unanswered. 
For example, Model Cards for Model Reporting~\cite{Mitchell2019} are validated through two hypothetical case studies; however, recent work has shown that the emphasis on accuracy-based metrics in Model Cards results in other downstream harms being overlooked, although the latter could be more beneficial for impacted communities~\cite{kawakami2024responsible}.
This finding is particularly critical, as it is mentioned in the documentation of Model Cards that they are meant to serve impacted communities. 
Overall, unvalidated RAI tools can be unusable, misused, or incompatible with existing practices, while among the few validated tools, many of the checks are removed from the real-world contexts where these tools are meant to be applied.

\xhdr{Fragmented approach to RAI} 
% Overemphasis on technical roles and stages could perpetuate a piecemeal approach to RAI, where solutions address isolated issues rather than systemic challenges. 
% This fragmented governance framework risks creating a disconnect between technical efforts and overarching ethical or organizational objectives, making it difficult to ensure coherence across the AI lifecycle. 
% Without a more holistic approach, systemic issues such as bias, opacity, or accountability gaps may persist, 
% % or misalignment with societal values may persist, 
% % undermining trust and the long-term sustainability of AI systems.
% continuing to make responsible AI difficult to realize in practice.
There are few validated RAI tools that address the AI lifecycle end-to-end 
% (Section~\ref{sec:lifecycle_stages}) 
and also address all the stakeholders that are involved in an AI system.
% (Section~\ref{sec:stakeholders}).
The lack of comprehensive and cohesive efforts for RAI means that many organizations are using a piecemeal approach to RAI and AI governance. 
Consequently, ethical issues at different stages of the AI lifecycle are being addressed in isolation.
However, this fragmented view of AI governance makes it difficult to have a holistic view of an AI system.
For example, AI Fairness 360~\cite{Bellamy2019} addresses fairness issues in data processing, statistical modeling, and testing.
However, ensuring fairness in these stages does not necessarily mean that fairness will translate into deployment. 
To illustrate, Cheng and Chouldechova~\cite{cheng2022heterogeneity} found that different end-users make different decisions with the same algorithmic output. 
In this example, while the algorithmic output may be fair, the end-user's decisions following the output may render the AI system unfair.
Approaching fairness in isolation can prevent organizations from understanding the interactions between all the different stages of the AI lifecycle where ethical issues can arise.

Furthermore, there is currently an absence of RAI tools that bring different stakeholders together. In particular, there are few tools that bring end-users and impacted communities together with other stakeholder groups.
Due to the wide-ranging impacts of AI systems, many people in RAI have advocated for involving, consulting with, and collaborating across diverse stakeholders~\cite{rakova2021responsible}.
However, there are few concrete tools to do so in the context of RAI, giving rise to difficulties in recognizing and reconciling different stakeholders' perspectives.
For instance, developers may benefit most from concrete, quantitative metrics for fairness, but impacted communities may benefit more from descriptions of potential downstream instances and implications of unfairness.
Without RAI tools bringing these two stakeholder groups together, developers may not understand impacted communities' needs, while impacted communities may have trouble understanding how an algorithm was tested for fairness and what the results were.

\section{Recommendations}
% Immediate Next Steps: Areas requiring urgent research focus, Suggestions for new tools tailored to underserved roles or lifecycle stages.\\
% Long-term Goals: Work with organizations to develop holistic approach
% \TODO{write a summary of the recommendations}
We provide three recommendations towards realizing RAI in practice: validating existing and new RAI tools, approaching AI governance in a holistic and end-to-end manner, and using the stakeholder-stage matrix (Figure~\ref{fig:roles_stages_sbs}) as a blueprint for RAI.
We elaborate on these below.
% To improve RAI tools and processes, researchers and practitioners must explore ways to validate tools in real world contexts while involving a broad range of stakeholders throughout the AI lifecycle to ensure that tools meet technical standards and societal needs. Additionally, research should be conducted to find ways to integrate RAI practices and tools throughout the AI lifecycle rather than retrofitted at the end. Also, researchers should find effective ways to foster collaboration between domain experts, technical professionals, and impacted community members. Future research should focus on exploring and creating a holistic blueprint  with flexible responsibility allocations for organizations. 

\xhdr{Validation of existing and new RAI tools} 
%%% Call for the validation of RAI tools
% Validation should be centered in real-world contexts
% Tools should be evaluated for effectiveness, rather than just usability
% In such evaluations, we look back to the stakeholder-stage matrix we used to classify RAI tools
% Like RAI tools themselves, evaluations of RAI tools can also benefit from this framework of thinking
% Echoing recent work surrounding RAI, RAI tools themselves should also be tested, validated, and monitored
% Resources for monitoring CITE
% They should also work with stakeholders to determine what effectiveness means
% The stakeholders that are meant to use the tool, and the stakeholders that are meant to use the tool’s output
% The exact mechanisms that it will take to achieve this in practice is still a nascent area. One idea is to borrow from some of the RAI tools we have explored, for instance, community jury
% What does it mean for an RAI tool targeted towards impacted communities to be effective?
%%% Recognition of the resource constraints of people that  are building RAI tools
% Organizational constraints
% Financial incentives for users needed for users to audit
% RAI stifles innovation
% However, we see this as an important step to ensure that we see RAI in practice
% In a world where we continue to have difficulties with seeing RAI in practice despite the growing interest in this area
% Through our systematic review of the literature, 
We discovered that most RAI tools lack any evidence of validation. For the few RAI tools that are validated, the validation method often fails to demonstrate whether the RAI tool will be effective in the real world.
Thus, we join previous calls~\cite{berman2024scoping} to validate RAI tools to ensure that they are effective in practice, that is, that using a given RAI tool allows real-world AI systems to achieve the claimed ethical values~\cite{cartwright2009thing}. 
We further recommend that RAI tool developers document both the validation process and results as part of the tool itself.

How can we validate RAI tools to ensure their effectiveness in the real world? 
Berman et al.~\cite{berman2024scoping} propose developing an ``effectiveness evaluation framework,'' outlining design desiderata for such a framework and providing guidelines for ensuring evaluation validity.
We argue that the stakeholder-stage matrix we used to classify RAI tools in this paper (Figure~\ref{fig:roles_stages_sbs}) is useful for thinking about the necessary components of RAI tool \emph{validations} as well.
To illustrate, RAI tools, like AI systems, should also go through stages such as testing (i.e. assessing the tool's effectiveness in driving desired changes in an organization's AI design, development, and deployment processes), validation (i.e. evaluating how well the tool generalized to different AI systems), and monitoring (i.e. ensuring the tool remains effective amidst evolving AI development practices).
% Note that there are other stages that should be considered in RAI tool validation; we use testing, validation, and monitoring as illustrative examples.
Furthermore, like AI systems, RAI tools validations should involve a variety of stakeholder groups. 
This involvement can help address specific outcomes that might otherwise be overlooked.
For example, organizational leaders could be involved to answer the question: what does it mean for an RAI tool to be effective for the organization? Similarly, impacted community members could be involved to answer the question: what does it mean for an RAI tool to be effective for their community?
% Similar to how previous research has recognized the need for multiple stakeholders to be involved in RAI efforts \TODO{because of xyz, Rachel cite}, we argue that the involvement of multiple stakeholders in RAI tool validations is critical for many of the same reasons.
This diverse stakeholder approach aligns with prior research emphasizing the importance of broad stakeholder involvement in RAI efforts~\cite{kaminski2018binary, kawakami2024responsible}. Similarly, we argue that involving multiple stakeholders in RAI tool validation is essential for ensuring their relevance and effectiveness. 

Achieving a multi-stage and multi-stakeholder RAI tool validation process presents significant challenges.
First, RAI tool developers have many resource constraints~\cite{berman2024scoping}.
Additionally, organizational practices and culture are often barriers to RAI~\cite{rakova2021responsible, varanasi2023currently}. Previous work has also shown that end-users require financial incentives to participate in auditing~\cite{deng2023understanding}
Furthermore, a narrow focus on validation can stifle innovation~\cite{greenberg2008usability}.
Nonetheless, in a world where we continue to have difficulties with seeing RAI in practice despite the growing interest in this area, we see RAI tool validations as a productive step forward.
As next steps, we recommend RAI tool developers to test, validate, and monitor the tools they create, consider multiple stakeholder groups, and document these processes within the RAI tool itself.
We also recommend researchers conduct further investigation into RAI tool evaluation practices, developing methods and frameworks to support effective validation.

\xhdr{A holistic, end-to-end approach to AI governance}
We discussed that the current landscape of RAI tools encourages an approach to RAI that is fragmented across both AI lifecycle stages and stakeholder groups, posing a barrier to realizing RAI in practice.
Thus, we join previous calls~\cite{Raji2020} to incorporate and develop an end-to-end approach to AI governance across lifecycle stages.
As illustrative examples of RAI tools that consider all stages of the AI lifecycle, we point to GAO's Accountability Framework~\cite{GAO2021} and Microsoft's AI Fairness Checklist~\cite{Madaio2020}.
We also recommend that approaches to AI governance be holistic not only across the AI lifecycle stages, but also across stakeholder groups.
For example, it may be beneficial to research architectures for collaborative lifecycle models in which designers, developers, and impacted communities jointly define the goals, constraints, and risks of AI systems.
In line with these recommendations, we call on organizational leaders to implement holistic processes within their workflows and researchers to focus their efforts on developing and testing the effectiveness of such processes.

\xhdr{A blueprint for RAI}
Ensuring strong accountability mechanisms is a difficult problem in RAI~\cite{cooper2022accountability, schiff2020principles}.
One of the reasons behind this difficulty is the ``many hands'' problem: there are many stages in the AI lifecycle, and many actors involved each stage~\cite{cooper2022accountability, nissenbaum1996accountability}
% The ``many hands'' problem makes 
Consequently, it is difficult to determine both who should be responsible for problems, but also where these problems stem from.

Our stakeholder-stage matrix (Figure~\ref{fig:roles_stages_sbs}) provides a structure not only for determining for each stakeholder what RAI tools are available when, but also for assigning clearer responsibilities for each of the stakeholders in an AI system across the AI lifecycle.
This is reminiscent of the responsibility assignment matrix, which is a widely-used project management tool encouraging an end-to-end view of a project and benefiting organizations with multiple moving parts~\cite{wittenberg2024everything}.
Indeed, one of the RAI tools in our literature review~\cite{Johnson2023} incorporates the responsibility assignment matrix to clearly delineate tasks to different stakeholder groups. 
Using a responsibility assignment matrix for AI systems can help disaggregate the extent of each stakeholder's involvement in each stage of the AI lifecycle and therefore serve as a potential remedy to the ``many hands'' problem.

\bkedit{We are currently working on a related project to understand how the blueprint can be used for AI assurance. To do so, we are interviewing members from each stakeholder group (leaders, designers, etc.) within real-world AI systems, and inquiring each stakeholder about their involvement in each stage of the AI lifecycle. 
This process is helping us explore which (stakeholder, stage)-pairs are responsible for the success or failure of the AI system. This in turn allows the organization to design a more effective allocation of responsibilities to different roles to operationalize responsible governance of AI.}

We encourage organizational leaders to consider the stakeholder-stage matrix as a blueprint for AI governance within their organization.
We also recommend that researchers explore the utility of the responsibility assignment matrix as a framework for RAI. 
For example, researchers could use the matrix to understand current organizational practices, determine whether responsibility allocations need to be adjusted depending on the specific AI use case, and establish the extent to which the responsibility assignment matrix is both standardized and customizable.

\bibliographystyle{ACM-Reference-Format}
\bibliography{refs}
\clearpage
\appendix

% \begin{landscape}
\section{Citation Count}\label{app:citation_count}
\begin{figure*}[]
    \centering
    \includegraphics[width=1\linewidth]{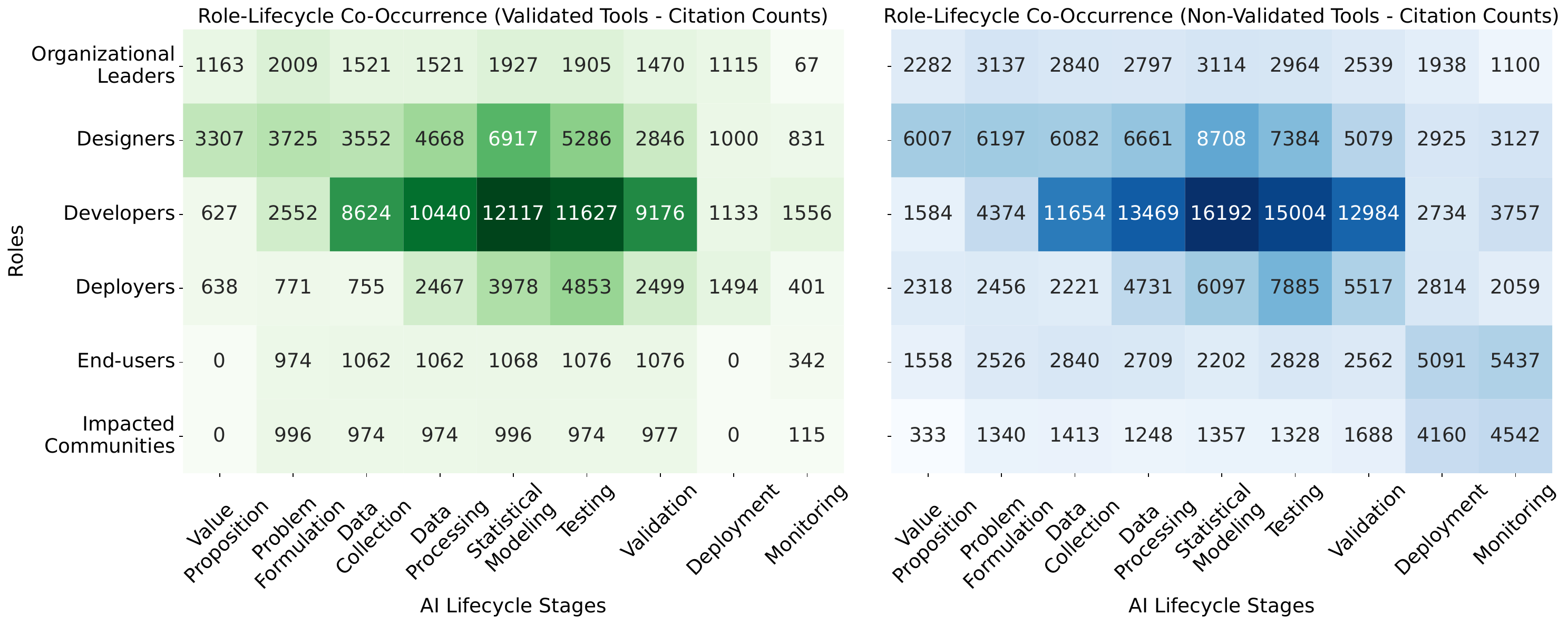}
    \caption{Citation counts for all validated tools \rkedit{(left)} and citation counts for validated and non-validated tools (\rkedit{right}) by \rkedit{stakeholder-stage} pairs. }
    \label{fig:citation_count}
\end{figure*}

\rkedit{Figure~\ref{fig:citation_count} is the co-occurrence matrix for stakeholders and stages using citation counts, rather than the number of tools.
We used the citation counts of the tools from Scopus and Google Scholar; however, we could not find the citation counts for all the tools, especially for tools from industry. When this was the case, the tool was not included in the matrix.
Overall, we see a similar trend to Figure~\ref{fig:roles_stages_sbs}: with regards to the stages, ``Data Collection'', ``Data Processing'', ``Statistical Modeling'', ``Testing'', and ``Validation'' are overrepresented, while the remaining stages are unrepresented.
With regards to the stakeholders, ``Developers'' are overrepresented, ``Designers'' and ``Deployers'' are moderately represented, and the remaining stakeholders are underrepresented.}

% \end{landscape}

\section{Tool Table}\label{app:tool_table}

\begin{table*}[]
\centering
\footnotesize
\caption{Validated Tools for each lifecycle stage and stakeholder role}
\begin{tabular}
{p{0.075\linewidth}p{0.075\linewidth}p{0.075\linewidth}p{0.075\linewidth}p{0.075\linewidth}p{0.075\linewidth}p{0.075\linewidth}p{0.075\linewidth}p{0.075\linewidth}p{0.075\linewidth}}
\toprule
 % & \rotatebox{45}{\textbf{Value Proposition}} 
 % & \rotatebox{45}{\textbf{Problem Formulation}} 
 % & \rotatebox{45}{\textbf{Data Collection}} 
 % & \rotatebox{45}{\textbf{Data Processing}} 
 % & \rotatebox{45}{\textbf{Statistical Modeling}} 
 % & \rotatebox{45}{\textbf{Testing}} 
 % & \rotatebox{45}{\textbf{Validation}} 
 % & \rotatebox{45}{\textbf{Deployment}} 
 % & \rotatebox{45}{\textbf{Monitoring}}\\ \midrule
  & \textbf{Value Proposition}
 & \textbf{Problem Formulation}
 & \textbf{Data Collection}
 & \textbf{Data Processing}
 & \textbf{Statistical Modeling}
 & \textbf{Testing}
 & \textbf{Validation}
 & \textbf{Deployment} 
 & \textbf{Monitoring}\\ \midrule
\textbf{Leaders}           &  \cite{Asatiani2021, GAO2021, Lobschat2021, Moon2019, Prem2023, Raji2020, Saetra2021, Vidgen2019} & \cite{Asatiani2021, El-Haber2024, GAO2021, Janssen2022, Lobschat2021, MS2022, Mitchell2019, Moon2019, Prem2023, Raji2020, Tyler2024}  & \cite{El-Haber2024, GAO2021, Hawkins2021, MS2022, Mitchell2019, Moon2019, Prem2023, Raji2020, Tyler2024}  & \cite{El-Haber2024, GAO2021, Hawkins2021, MS2022, Mitchell2019, Moon2019, Prem2023, Raji2020, Tyler2024}  &  \cite{El-Haber2024, GAO2021, Hawkins2021, Janssen2022, MS2022, Mitchell2019, Moon2019, Prem2023, Raji2020, Saleiro2018, Tyler2024} & \cite{El-Haber2024, GAO2021, Hawkins2021, Mitchell2019, Moon2019, Prem2023, Raji2020, Saleiro2018, Tyler2024}  &  \cite{El-Haber2024, GAO2021, Hawkins2021, MS2022, Mitchell2019, Moon2019, Prem2023, Raji2020, Tyler2024} &  \cite{El-Haber2024, GAO2021, Hawkins2021, Lobschat2021, Moon2019, Prem2023, Saleiro2018, Tyler2024} & \cite{GAO2021, Prem2023, Saetra2021, Sattlegger2024, Tyler2024}  \\ \midrule
\textbf{Designers}           &  \cite{Asatiani2021, EUAI2020, Economou2023, Friedman2013, GAO2021, Madaio2020, Moon2019, Nelson2020, Purzer2024, Raji2020, Vidgen2019} & \cite{Adkins2022, Arnold2018, Asatiani2021, Dreossi2019, EUAI2020, Economou2023, El-Haber2024, Friedman2013, GAO2021, Janssen2022, Lane2018, MS2022, Madaio2020, Moon2019, Naja2022, Nelson2020, Purzer2024, Raji2020, TF2025, Tyler2024}  & \cite{Adkins2022, Ahn2020, Bender2018, Corra2021, DCPB2021, Economou2023, El-Haber2024, GAO2021, Holland2018, Hsu2021, Kroll2017, Lo2023, MS2022, Madaio2020, Moon2019, Naja2022, Nelson2020, Purzer2024, Raji2020, Singh2019, TF2025, TW2023, Tramer2017, Tyler2024, UK2023, aifs360-2019}  & \cite{Adkins2022, Ahn2020, Bantilan2017, Bellamy2019, Corra2021, DCPB2021, Dreossi2019, Economou2023, El-Haber2024, GAO2021, Holland2018, Hsu2021, Kroll2017, Lo2023, MS2022, Madaio2020, Moon2019, Naja2022, Nelson2020, Raji2020, Singh2019, TF2025, TW2023, Tramer2017, Tyler2024, UK2023, aifs360-2019}  &  \cite{Adkins2022, Ahn2020, Bantilan2017, Bellamy2019, Bender2018, Corra2021, Dreossi2019, Economou2023, El-Haber2024, GAO2021, Holland2018, Janssen2022, Lo2023, MS2022, Madaio2020, Moon2019, Naja2022, Nelson2020, Purzer2024, Raji2020, Saleiro2018, Singh2019, TW2023, Tramer2017, Tyler2024, Weerts2023, Wexler2020, aifs360-2019} & \cite{Adkins2022, Ahn2020, Bantilan2017, Bellamy2019, Corra2021, Dreossi2019, El-Haber2024, GAO2021, Madaio2020, Moon2019, Naja2022, Nelson2020, Nushi2018, Raji2020, Saleiro2018, Tramer2017, Tyler2024, Weerts2023, Wexler2020, aifs360-2019}  &  \cite{Adkins2022, Ahn2020, Bantilan2017, Dreossi2019, Economou2023, El-Haber2024, GAO2021, Kroll2017, Lo2023, MS2022, Madaio2020, Matias2022, Moon2019, Naja2022, Nelson2020, Nushi2018, Raji2020, Tramer2017, Tyler2024, Weerts2023, aifs360-2019} &  \cite{Contractor2022, Corra2021, Economou2023, El-Haber2024, GAO2021, Lo2023, Madaio2020, Moon2019, Nelson2020, PDPC2020, Saleiro2018, TF2025, Tyler2024, UK2023, Weerts2023} & \cite{Corra2021, GAO2021, Kroll2017, Lane2018, Lo2023, Madaio2020, Singh2019, Tramer2017, Tyler2024, UK2023}  \\ \midrule
\textbf{Developers}          &  \cite{Asatiani2021, EUAI2020, Economou2023, Madaio2020, Moon2019, Prem2023, Purzer2024, Raji2020, Saetra2021} & \cite{Adkins2022, Arnold2019, Asatiani2021, Ballard2019, Deon2023, Dreossi2019, EUAI2020, Economou2023, El-Haber2024, Janssen2022, MS2022, Madaio2020, Mitchell2019, Moon2019, Naja2022, Prem2023, Purzer2024, Raji2020, TF2025, Tyler2024}  & \cite{Adkins2022, Ahn2020, Arnold2019, Ballard2019, Bates2019, Bender2018, Corra2021, DCPB2021, Deon2023, Economou2023, El-Haber2024, Gebru2021, Hawkins2021, Holland2018, Hsu2021, Kroll2017, Lam2024, Landers2022, Lo2023, MS2022, Madaio2020, Mitchell2019, Moon2019, Naja2022, Prem2023, Purzer2024, Raji2020, Suresh2022, TF2025, TFF2024, Tramer2017, Tyler2024, Vasudevan2020, Wilson2021, Yu2019, aifs360-2019, dAlessandro2017}  & \cite{Adkins2022, Ahn2020, Arnold2019, Bantilan2017, Bates2019, Bellamy2019, Corra2021, DCPB2021, Deon2023, Dreossi2019, Economou2023, El-Haber2024, Gebru2021, Hall2021, Hawkins2021, Holland2018, Hsu2021, Kroll2017, Landers2022, Lo2023, MS2022, Madaio2020, Mitchell2019, Moon2019, Naja2022, Prem2023, Raji2020, Ryffel2018, Suresh2022, TF2025, TFF2024, Tramer2017, Tyler2024, Vasudevan2020, Whang2021, Wilson2021, Wiśniewski2022, Yu2019, aifs360-2019, dAlessandro2017}  &  \cite{Adkins2022, Ahn2020, Arnold2019, Aronsson2021, Bantilan2017, Bellamy2019, Bender2018, Corra2021, Deon2023, Dreossi2019, Economou2023, El-Haber2024, Goldstein2013, Hall2021, Hawkins2021, Holland2018, Janssen2022, Landers2022, Lo2023, MS2022, Madaio2020, Mitchell2019, Moon2019, Naja2022, Panigutti2021, Prem2023, Purzer2024, Raji2020, Ryffel2018, Saleiro2018, Suresh2022, TFF2024, Tenney2020, Tramer2015, Tramer2017, Tyler2024, Vasudevan2020, Viswanath2023, Weerts2023, Wexler2020, Whang2021, Wilson2021, Wiśniewski2022, aifs360-2019, dAlessandro2017} & \cite{Adkins2022, Ahn2020, DAmour2020, Arnold2019, Bantilan2017, Bellamy2019, Corra2021, Deon2023, Dreossi2019, El-Haber2024, Friedler2019, Gebru2021, Hawkins2021, Landers2022, Madaio2020, Mitchell2019, Moon2019, Naja2022, Nushi2018, Panigutti2021, Prem2023, Raji2020, Saleiro2018, Suresh2022, TFF2024, Tenney2020, Tramer2015, Tramer2017, Tyler2024, Vasudevan2020, Viswanath2023, Weerts2023, Wexler2020, Wiśniewski2022, aifs360-2019, dAlessandro2017}  &  \cite{Adkins2022, Ahn2020, DAmour2020, Arnold2019, Bantilan2017, Deon2023, Dreossi2019, Economou2023, El-Haber2024, Epstein2018, Friedler2019, Gebru2021, Hawkins2021, Kroll2017, Lam2024, Landers2022, Lo2023, MS2022, Madaio2020, Mitchell2019, Moon2019, Naja2022, Nushi2018, Prem2023, Raji2020, Suresh2022, TFF2024, Tenney2020, Tramer2015, Tramer2017, Tyler2024, Vasudevan2020, Viswanath2023, Weerts2023, Wilson2021, Wiśniewski2022, aifs360-2019, dAlessandro2017} &  \cite{Corra2021, Deon2023, Economou2023, El-Haber2024, Hawkins2021, IBM2024, Lo2023, Madaio2020, Moon2019, PDPC2020, Prem2023, Saleiro2018, TF2025, Tyler2024, Weerts2023} & \cite{Arnold2019, Corra2021, Deon2023, Epstein2018, IBM2024, Kroll2017, Liu2022, Lo2023, Madaio2020, Prem2023, Saetra2021, Tramer2017, Tyler2024, Wilson2021}  \\ \midrule
\textbf{Deployers}           &  \cite{EUAI2020, GAO2021, Lobschat2021, Madaio2020, Moon2019, Prem2023} & \cite{Ballard2019, EUAI2020, GAO2021, Lobschat2021, Madaio2020, Moon2019, Prem2023, TF2025}  & \cite{Ahn2020, Ballard2019, Corra2021, GAO2021, Lam2024, Madaio2020, Moon2019, Prem2023, TF2025, TFF2024, Tramer2017, Vasudevan2020, Yu2019}  & \cite{Ahn2020, Bellamy2019, Corra2021, GAO2021, Madaio2020, Moon2019, Prem2023, TF2025, TFF2024, Tramer2017, Vasudevan2020, Yu2019}  &  \cite{Adler2018, Ahn2020, Aronsson2021, Bellamy2019, Corra2021, GAO2021, Madaio2020, Moon2019, Panigutti2021, Prem2023, Saleiro2018, TFF2024, Tenney2020, Tramer2015, Tramer2017, Vasudevan2020, Viswanath2023, Weerts2023} & \cite{Adler2018, Ahn2020, DAmour2020, Bellamy2019, Corra2021, Friedler2019, GAO2021, Madaio2020, Moon2019, Munoz2024, Panigutti2021, Prem2023, Saleiro2018, TFF2024, Tenney2020, Tramer2015, Tramer2017, Vasudevan2020, Viswanath2023, Weerts2023}  &  \cite{Adler2018, Ahn2020, DAmour2020, Friedler2019, GAO2021, Lam2024, Madaio2020, Matias2022, Moon2019, Munoz2024, Prem2023, TFF2024, Tenney2020, Tramer2015, Tramer2017, Vasudevan2020, Viswanath2023, Weerts2023} &  \cite{Corra2021, GAO2021, Lobschat2021, Madaio2020, Moon2019, PDPC2020, Prem2023, Saleiro2018, TF2025, Weerts2023} & \cite{Adler2018, Corra2021, GAO2021, Madaio2020, Prem2023, Tramer2017}  \\ \midrule
\textbf{End-users}           &  [] & \cite{El-Haber2024, Mitchell2019}  & \cite{El-Haber2024, Mitchell2019, Singh2019, Suresh2022}  & \cite{El-Haber2024, Mitchell2019, Singh2019, Suresh2022}  &  \cite{Aronsson2021, El-Haber2024, Mitchell2019, Singh2019, Suresh2022} & \cite{El-Haber2024, Mitchell2019, Nushi2018, Suresh2022}  &  \cite{El-Haber2024, Mitchell2019, Nushi2018, Suresh2022} &  \cite{El-Haber2024} & \cite{Cabrera2021, Liu2022, Singh2019, Song2019}  \\ \midrule
\textbf{Impacted Communities} &  \cite{MSCJ2022} & \cite{El-Haber2024, Janssen2022, Lane2018, MSCJ2022, Mitchell2019}  & \cite{El-Haber2024, Mitchell2019}  & \cite{El-Haber2024, Mitchell2019}  &  \cite{El-Haber2024, Janssen2022, Mitchell2019} & \cite{El-Haber2024, Mitchell2019}  &  \cite{El-Haber2024, Matias2022, Mitchell2019} &  \cite{El-Haber2024} & \cite{Lane2018, Song2019}  \\ \bottomrule
\end{tabular}
\label{tab:validated_tools}
\end{table*}
% \end{landscape}

% \subsection{Non-Validated Tool Table}

% \clearpage

\end{document}